\documentclass[a4paper, amsfonts, amssymb, amsmath, twoside, onecolumn, showkeys, nofootinbib, aps, pra, 10pt]{revtex4-2}
\usepackage[english]{babel}
\usepackage[utf8]{inputenc}
\usepackage[colorinlistoftodos, color=green!40, prependcaption]{todonotes}
\usepackage[pdftex, pdftitle={Article}, pdfauthor={Author}]{hyperref}
\usepackage{amsthm}
\usepackage{mathtools}
\usepackage{dsfont}
\usepackage{physics}
\usepackage{xcolor}
\usepackage{graphicx}
\usepackage[paperwidth=210mm, paperheight=297mm, centering, hmargin=2cm, vmargin=2.5cm]{geometry}
\usepackage{adjustbox}
\usepackage{placeins}
\usepackage[T1]{fontenc}
\usepackage{lipsum}
\usepackage{csquotes}
\usepackage{float}
\usepackage{comment}

\begin{document}
\title{Squeezed states for Frenkel-like two-fermion composite bosons}

\author{Francisco Figueiredo}
    \email[Correspondence email address: ]{frfig@cbpf.br}
    \affiliation{Centro Brasileiro de Pesquisas Físicas, Rua Dr. Xavier Sigaud 150, 22290-180 Rio de Janeiro, Brazil}

\author{Itzhak Roditi}
    \email[Correspondence email address: ]{roditi@cbpf.br}
    \affiliation{Centro Brasileiro de Pesquisas Físicas, Rua Dr. Xavier Sigaud 150, 22290-180 Rio de Janeiro, Brazil}

\date{\today} 

\begin{abstract}

We investigate squeezed states of composite bosons (cobosons) formed by pairs of spin-$1/2$ fermions, with emphasis on Frenkel-like cobosons. We define squeezed cobosons as eigenstates of a Bogoliubov-transformed coboson operator and derive explicit expressions for the associated quadrature variances. We show that the underlying fermionic structure leads to state-dependent modifications of the Heisenberg--Robertson uncertainty bound, which may fall below the canonical bosonic limit without implying any violation of uncertainty principles. Numerical results based on finite-dimensional matrix representations illustrate how these effects constrain the attainable squeezing.
This behavior implies that experimentally accessible quantities, such as quadrature variances or noise spectra, carry information about the internal fermionic structure of the composite particles. Our framework is relevant to composite-boson systems such as tightly bound electron-hole pairs and provides a physically transparent setting to probe compositeness through observable quadrature fluctuations.

\end{abstract}


\maketitle

\section{Introduction} \label{sec:introduction}

In quantum optics, squeezed states are states that minimize the product of quadratures variances (uncertainties), for instance, $\Delta X \Delta P$, as also do Glauber coherent states (for short, Glauber states \cite{GlauberStates}); however, squeezed states allow one of the uncertainties to be arbitrarily low, within the bound of the Heisenberg uncertainty relation, in exchange for the increase of the other. One of the most striking current uses of squeezed states is in gravitational wave detection with laser interferometry (for a review, see \cite{SCHNABEL20171}). Most of the works investigating Glauber states and squeezed states are performed naturally with light, but the concept has also been extended to other degrees of freedom, such as, for example, spin squeezing \cite{PhysRevA.47.5138} which was recently implemented in an NMR setup \cite{PhysRevLett.114.043604} and deformed oscillators \cite{RAMIREZ20161117}.

In this work, our interest is to build squeezed states for the case of composite bosons, cobosons for short. Cobosons are states formed by pairs of fermions with opposite spins. The coboson formalism emerged in the early 2000s with the aim of understanding the approximate bosonic behavior of many-exciton systems \cite{Combescot2001, Combescot2003}, while also providing a theoretical framework that fully accounts for the fermionic nature of their constituents, in contrast with approaches based on bosonization procedures \cite{combescot2002effective, combescot2005many, EZAWA1998223}. Moreover, this formalism can be generalized to the study of arbitrary composite particles \cite{combescot2010general} and extended to scenarios involving finite temperature \cite{CobosonZ_Shiau2015, combescot2011general_finite_T}. There are also interesting connections of the composite boson formalism with deformed oscillators \cite{GAVRILIK20121596}.

In order to pursue this idea and investigate squeezed states, we use an approach similar to that of \cite{Combescot_coherent,SHIAU2023169431} where the construction of coboson Glauber states is attempted. For this, one uses an analogy with Frenkel excitons \cite{FrenkelPhysRev.37.17}, these excitons are constituted by tightly bound electron-hole pairs where the electron and the hole are on the same site in a lattice, nevertheless they are delocalized via intersite interatomic-level Coulomb processes \cite{CombescotExcitonsCooperPairs}. The treatment in the case of Frenkel-like cobosons is simplified by the fact that the Schmidt decomposition of the wave function encoding the two species of fermions follows a flat distribution of its components. 

The extension to cobosons has more intricate algebraic properties which is related to the presence of fermions and leads, for instance, to Pauli-blocking effects \cite{Combescot_coherent} and non-trivial commutation rules. These corrections will appear when studying the quadratures as well as the Bogoliubov transformation. In addition to formal interests, we believe that the notion of squeezing developed here for cobosons is relevant for those systems involving composite bosons, as in exciton polaritons \cite{ZHANG202029}, quasi-particles obtained by coupling semiconductor excitons and cavity photons. So, understanding squeezing in cobosons (which can also be formed by pairing of bosonic modes) may turn out to be useful in realistic systems used in quantum control or precision measurements.

This work is structured as follows: In the next Section, we outline the coboson formalism centering around Frenkel-like cobosons. In Section III we present our construction of squeezed Frenkel-like cobosons that are defined as the eigenstates of a Bogoliubov transformed coboson operator. In Section IV we discuss properties, and in Section V we present some numerical results for the obtained squeezed states. Our concluding remarks are contained in Section VI.

\section{Outline of the formalism for Frenkel-like cobosons} \label{sec:formalism}

The fundamental building blocks of the formalism are states of fermions of two species encoded in a wavefunction $\Psi,$ where one obtains its Schmidt decomposition \cite{Law2005,Bouvrie-Majtey-Figueiredo-Roditi_2019},  
\begin{equation}
\ket{\Psi}= \sum_{k=1}^S \sqrt{\lambda_k} \ket{a_k}\ket{ b_k} ,  \label{SchmidtDecomp}\;\;
 \lambda_1 \ge \lambda_2 \ge \dots  \ge 0, ~~ \sum_{k=1}^S \lambda_k =1,   
\end{equation}
 $\ket{a_k}$ and $\ket{b_k},$ are two single-particle bases.

One then treats a fermion pair as a coboson with an associated creation operator (using the corresponding Schmidt decomposition)  given by 
\begin{align}
    B^\dagger= \sum\limits_{k}^S\sqrt{\lambda_k }a^\dagger_{k} b^\dagger_{k},
\end{align}
the different modes follow fermionic canonical commutation rules.

Frenkel-like cobosons are those whose Schmidt decomposition is flat \cite{Combescot_coherent}, meaning that their decomposition coefficient, $\lambda_{k}$, is a phase that gives a constant $|\lambda_{k}|$, similar to the behavior of Frenkel excitons \cite{Combescot2008Frenkel}. The usual creation and destruction operators commutation rules in the cobosons formalism are given by,
\begin{align}\label{CommRules}
    [B, B^\dagger]= 1 - D.
\end{align}

The operator $D$ is given by a positive combination of $a^{\dagger}_la_l$ and $b^{\dagger}_lb_l$, it is thus a positive operator. This means that one can attain uncertainty properties that are smaller than the, usual $1/2$, without violating the Heisenberg limit, because the commutation rule Eq.\ref{CommRules} deviates from the canonical one.
For Frenkel-like cobosons, the coefficients are simply a phase, we then have,
\begin{align}
    B^\dagger = \frac{1}{\sqrt{N_s}}\sum\limits_{k=1}^{N_s}e^{i\theta_k} a^\dagger_k b^\dagger_k,
\end{align}
where $N_s$ is the number of pair states needed to build the respective coboson.
In that case, the commutation relations with the operator $D$ have the form,
\begin{align}
    &[D, B^\dagger]= \frac{2}{N_s} B^\dagger,\\\nonumber
    &[D, B]= -\frac{2}{N_s} B.
\end{align}

It is possible to obtain Fock states \cite{Law2005,Combescot_coherent} such that for,
\begin{align}
    \ket{N} = \frac{1}{\sqrt{N!\,\chi_N}}\left( B^\dagger\right)^N\ket{0},
\end{align}

where $\chi_{N} = \frac{N_s!}{
N_s^N (N_s - N)!},$ and the coboson operators act as
\begin{align}\label{fock1}
    & B^\dagger\ket{N} = F_{N+1}\ket{N+1}\\\nonumber
    & B \ket{N} = F_N\ket{N-1}
\end{align}
with $F_N = \sqrt{N\left( 1-\frac{N-1}{N_s}.
\right)}.$

\section{Squeezed state for Frenkel-like cobosons} \label{sec:squeezed}

Here we deal with squeezed states; analogously to the usual bosonic case \cite{Gilbert2010IQO, agarwal2013QO}, we define them as the eigenstates of the operator obtained via a Bogoliubov transformation of the creation and destruction operators. This approach has some similarity with that of \cite{Combescot_coherent} where quasi-Glauber states are defined, in their seminal study they had an upper limit of $N_s-1$, for the eigenstates of $B$, which prevented them of having Glauber states. In what follows, we do not have that limitation.

In our framework a Bogoliubov transformed coboson operator is given by;
\begin{align}
    \mathcal{B}_\xi = \cosh{r}\, B+ e^{i\varphi}\sinh{r}\, B^\dagger,
\end{align}
where $\xi=r\, e^{i\varphi}.$ We call $r$ a squeezing parameter.\\

One can easily check that such a transformation preserves the usual coboson commutation relation.

We, therefore, define the coboson squeezed state $\ket{\alpha,\xi}$ as 
\begin{align}
    \mathcal{B}_\xi\ket{\alpha,\xi} = \alpha \ket{\alpha,\xi}
\end{align}

Essentially, what we want is a state $\sum\limits_{N=0}^{N_s}x_N \ket{N}$, satisfying 
\begin{align}
    \mathcal{B}_\xi \left( \sum\limits_{N=0}^{N_s}x_N \ket{N} \right) = \alpha \sum\limits_{N=0}^{N_s}x_N \ket{N}.
\end{align}

From that we have

\begin{align}
    &\mathcal{B}_\xi \left( \sum\limits_{N=0}^{N_s}x_N \ket{N} \right) =  \\ \nonumber
    &= \cosh{r}\, \sum\limits_{N=1}^{N_s}x_{N}F_N\ket{N-1}\\\nonumber &\qquad + e^{i\varphi}\sinh{r}\, \sum\limits_{N=0}^{N_s-1}x_{N}F_{N+1}\ket{N+1} \\\nonumber
    &= \cosh{r}\, \sum\limits_{N=0}^{N_s-1}x_{N+1}F_{N+1}\ket{N}\\\nonumber &\qquad + e^{i\varphi}\sinh{r}\, \sum\limits_{N=1}^{N_s}x_{N-1}F_{N}\ket{N}\\ \nonumber
    &= \alpha \sum\limits_{N=0}^{N_s}x_N \ket{N}. \nonumber
\end{align}
Then, the above eigenvalue condition leads to a system of equations that can be cast in matrix form as,
\begin{equation}  
\begin{bmatrix}
0 & C_1 & 0 & 0 & \cdots & 0 \\
S_1 & 0 & C_2 & 0 & \cdots & 0 \\
0 & S_2 & 0 & C_3 & \cdots & 0 \\
0 & 0 & S_3 & 0 & \cdots & \vdots \\
\vdots & \vdots & \vdots & \vdots & \ddots & C_{N_s} \\
0 & 0 & 0 & \cdots & S_{N_s} & 0
\end{bmatrix}
\begin{bmatrix}
x_0 \\ x_1 \\ x_2 \\ \vdots \\ x_{N_{S-1}} \\  x_{N_{S}}
\end{bmatrix}
=
\alpha
\begin{bmatrix}
x_0 \\ x_1 \\ x_2 \\ \vdots \\ x_{N_{S-1}} \\  x_{N_{S}}
\end{bmatrix}
,\label{tri}\end{equation} where $C_N = \cosh(r) F_N$, and $S_N = e^{i\varphi}\sinh (r)F_N .$ 

It is interesting to observe the fact that the eigenvalues of the tri-diagonal matrix (with $0$'s in the diagonal) in Eq.\ref{tri}, come in $\pm$pairs for even dimensions and, in addition,  must have a $0$ eigenvalue for odd dimensions (see, for instance, \cite{dyachenko2022tridiagonal}). Usually, the squeezed state with null eigenvalue is called a squeezed vacuum.

\section{Properties of the squeezed states} \label{sec:properties}

Here we will investigate the coboson quadrature variances, in this direction inspired by the position and momentum of the harmonic oscillator, we define the quadrature operators,
\begin{align}
    &\hat{\chi} = \frac{B+B^{\dagger}}{\sqrt{2}} \\
    &\hat{\pi} = \frac{B-B^{\dagger}}{\sqrt{2}\mathrm{i}} .
\end{align}
The coboson quadrature variance for $\hat{\chi}$ is then
\begin{align}
    &\Delta\hat{\chi} = \sqrt{ \langle\hat{\chi}^2\rangle - \langle\hat{\chi}\rangle^2} \\\nonumber
   &\!=\! \frac{1}{\sqrt{2}}\left\{ \left\langle B^2 + {B^{\dagger}}^2 + BB^{\dagger} + B^{\dagger}B\right\rangle \!-\! \left\langle \!B+ B^{\dagger}\right\rangle^2 \right\}^{\frac{1}{2}}.
\end{align}

The Heisenberg-Robertson bound \cite{PhysRev.34.163} is state dependent and involves the expectation value of the commutator, 
\begin{align}\label{HR-bound}
  \Delta\hat{\chi}\Delta\hat{\pi}\geq \frac{1}{2}\vert \langle \comm{\hat{\chi}}{\hat{\pi}} \rangle  \vert.
\end{align}
In our case,
\begin{align}\label{XPComm}
\comm{\hat{\chi}}{\hat{\pi}} = i (1-D),
\end{align}
as $D$ is semi-positive definite, the right hand-side of Eq.\ref{HR-bound} may be smaller than the canonical $1/2,$ and is not a violation of the bound.

One hallmark of squeezed states is that, within the Heisenberg-Robertson bound, one of the quadrature variances, say $\Delta\hat{\chi}$, can be lowered in exchange for the increase of the value of the other one, in our case,  $\Delta\hat{\pi}$. Notice that one can write both $\hat{\chi}$ and $\hat{\pi}$ in terms of the Bogoliubov transformed $B_\xi$, as, for $\varphi=0$,
\begin{align}
    &B = \mathrm{cosh}(r) B_\xi - \mathrm{sinh}(r) B_\xi^\dagger \\\nonumber
    &B^\dagger = -\mathrm{sinh}(r) B_\xi + \mathrm{cosh}(r) B_\xi^\dagger ,\label{inverseB}
\end{align}
such that
\begin{align}
    \hat{\chi} &= \frac{B+B^\dagger}{\sqrt{2}}=\frac{\mathrm{e}^{-r}}{\sqrt{2}}(B_\xi+B_\xi^\dagger)\\\nonumber
    \hat{\pi} &= \frac{B-B^\dagger}{\sqrt{2}\mathrm{i}}=\frac{\mathrm{e}^{+r}}{\mathrm{i}\sqrt{2}}(B_\xi-B_\xi^\dagger).\\\label{ChiPiB}
\end{align}
From this, one has
\begin{align}
    \hat{\chi}^2 = \frac{\mathrm{e}^{-2r}}{2}(B_\xi^2+B_\xi^{\dagger\, 2} + 2  B_\xi^\dagger B_\xi +1 - D).
\end{align}
Taking the average $\langle \hat{\chi}^2 \rangle_\alpha\equiv  \langle\alpha\vert \hat{\chi}^2 \vert \alpha \rangle,$ over a squeezed state $B_\xi\vert\alpha\rangle = \alpha \vert\alpha\rangle,$ and denoting $d \equiv \langle D \rangle_\alpha$, we arrive at the following,
\begin{align}
    \langle \hat{\chi}^2 \rangle_\alpha=\frac{\mathrm{e}^{-2r}}{2} ( (\alpha + \alpha^*)^2 + 1 -d ).
\end{align}
with this result,
\begin{align}
    \Delta \hat{\chi}^{\,2} = \langle\chi^2\rangle_\alpha - \langle\chi\rangle_\alpha^2 = \frac{\mathrm{e}^{-2r}}{2} (1-d).
\end{align}
Analogously,
\begin{align}
    \Delta \hat{\pi}^{\,2} = \langle\pi^2\rangle_\alpha - \langle\pi\rangle_\alpha^2 = \frac{\mathrm{e}^{+2r}}{2} (1-d).
\end{align}

For a usual bosonic squeezed state (i.e. $d=0$) one recovers the standard result, see, for instance, \cite{agarwal2013QO}. If $d \rightarrow 1 $ the variance approaches zero, indicating a suppression of the fluctuations. It is important to keep in mind that physically one always has a finite number of pairs such that $d$ is bounded, and this kind of limit requires care.

\subsection*{Matrices for $\hat{\chi}$ and $\hat{\pi}$}

It is possible to write the coboson quadratures $\hat{\chi}$ and $\hat{\pi}$, for a given number of pair states, $N_s$, as matrices.

\begin{align}
    &\hat{\chi} = \frac{1}{\sqrt{2}}
    \begin{bmatrix}
        0 & F_1 & 0 & 0 & \cdots & 0 \\
        F_1 & 0 & F_2 & 0 & \cdots & 0 \\
        0 & F_2 & 0 & F_3 & \cdots & 0 \\
        0 & 0 & F_3 & 0 & \cdots & \vdots \\
        \vdots & \vdots & \vdots & \vdots & \ddots & F_{N_s} \\
        0 & 0 & 0 & \cdots & F_{N_s} & 0
    \end{bmatrix}\\
    &\hat{\pi} = \frac{1}{\sqrt{2}i}
    \begin{bmatrix}
        0 & F_1 & 0 & 0 & \cdots & 0 \\
        -F_1 & 0 & F_2 & 0 & \cdots & 0 \\
        0 & -F_2 & 0 & F_3 & \cdots & 0 \\
        0 & 0 & -F_3 & 0 & \cdots & \vdots \\
        \vdots & \vdots & \vdots & \vdots & \ddots & F_{N_s} \\
        0 & 0 & 0 & \cdots & -F_{N_s} & 0
    \end{bmatrix}  
\end{align}
With the above expressions, the typical quantities to be calculated are obtained by simple linear algebra operations, and can be established numerically by standard tools as Mathematica or Python. For example, letting $x$ be a column matrix representing one of the  eigenvectors obtained in Eq.\ref{tri} (i.e. a squeezed state), we can write 
\begin{align}
    &\langle\hat{\chi}\rangle = x^\dagger \hat{\chi} x \\ \nonumber
    &\langle\hat{\chi^2}\rangle = x^\dagger \hat{\chi}^2 x. \label{linearoperations}
\end{align}
One has analogous expressions also for the $\hat{\pi}$ observable.

\section{Results} \label{sec:results}

In this section, we present some numerical results of the quadrature variances
$\Delta\hat{\chi}$ and $\Delta\hat{\pi}$, focusing on how squeezing is modified by the composite nature of cobosons. The numerical results are obtained from the finite-dimensional matrix representations of the quadrature operators introduced in Sect.~\ref{sec:properties}. Specifically, expectation values and variances are computed using the eigenvectors of the tridiagonal matrices associated with the Bogoliubov-transformed coboson operators. These eigenvectors are labeled by an index running from $1$ to $N_s$, where $N_s$ is the number of fermion pairs; throughout this section, we focus on the eigenstate with index $N_s$, which is the most sensitive to finite-size and Pauli blocking effects. Since the value of the phase $\varphi$ does not qualitatively affect the results, we set $\varphi=0$ for simplicity.

\begin{figure}
\centering
\fcolorbox{blue}{white}{\includegraphics[width=0.45\textwidth]{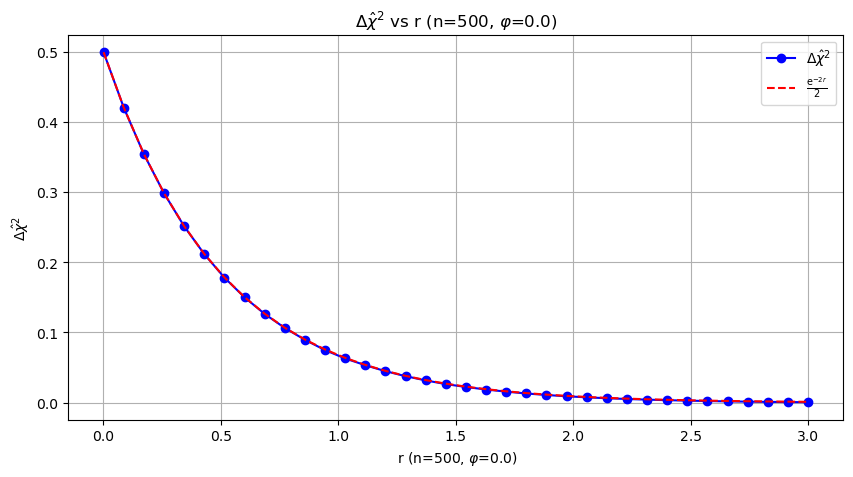}}
\caption{$(\Delta\hat{\chi})^2$ calculated for the eigenstate of the squeeze operator indexed by $N_s=n$, and different values of $r$. Those values are superposed to the value of the uncertainties of this quadrature for usual bosonic modes, $\frac{\mathrm{e}^{-2r}}{2}$ \cite{agarwal2013QO}. For $(\Delta\hat{\chi})^2$ the fact that the exponential decreases very fast gives a seemingly good agreement.}
\label{fig:framed_plot1}
\end{figure}

Figure~\ref{fig:framed_plot1} shows the behavior of $(\Delta\hat{\chi})^2$ as a function of the squeezing parameter $r$. The numerical results are compared with the canonical bosonic prediction $(\Delta\hat{\chi})^2 = e^{-2r}/2$  \cite{agarwal2013QO}.
Because this variance decreases exponentially with $r$, deviations due to compositeness remain relatively small throughout the explored range, leading to an apparent agreement with the elementary bosonic case. This behavior reflects the fact that squeezing in this quadrature suppresses fluctuations, which, in this case, are less sensitive to the finite-dimensional nature of the coboson Hilbert space.

\begin{figure}
\centering
\fcolorbox{blue}{white}{\includegraphics[width=0.45\textwidth]{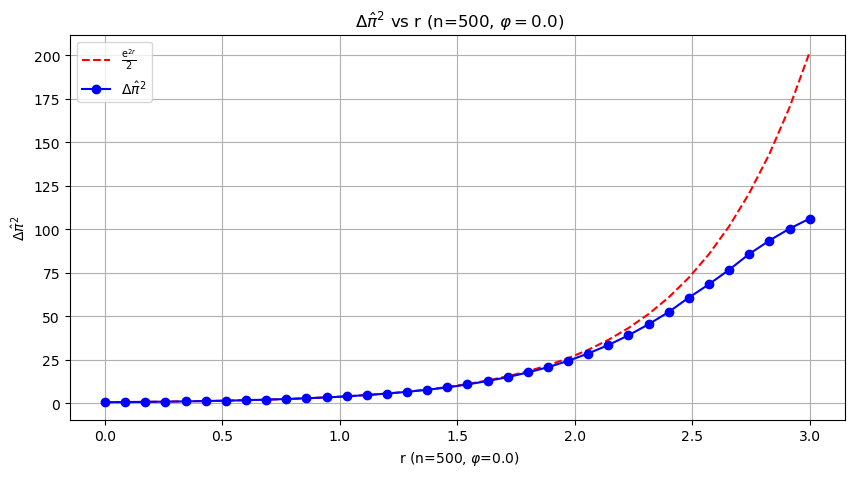}}
\caption{$(\Delta\hat{\pi})^2$ calculated for the eigenstate of the squeeze operator indexed by $N_s=n$, and different values of $r$. Those values are superposed to the value of the uncertainties of this quadrature for usual bosonic modes, $\frac{\mathrm{e}^{2r}}{2}$ \cite{agarwal2013QO}. For $(\Delta\hat{\pi})^2$ the fact that the exponential increases very fast gives a picture of the deviation from the usual bosonic behavior.}
\label{fig:framed_plot2}
\end{figure}

In contrast, Fig.~\ref{fig:framed_plot2} displays $(\Delta\hat{\pi})^2$ as a function of $r$, together with the bosonic reference $(\Delta\hat{\pi})^2 = e^{2r}/2$. Here, deviations from the canonical behavior become pronounced at large squeezing. The exponential amplification of fluctuations in the conjugate quadrature competes with the finite number of available fermion pairs, leading to a clear saturation effect that has no analog in the elementary bosonic case. This behavior provides a direct numerical manifestation of Pauli blocking and finite pair occupancy.

\begin{figure}
\centering
\fcolorbox{blue}{white}{\includegraphics[width=0.46\textwidth]{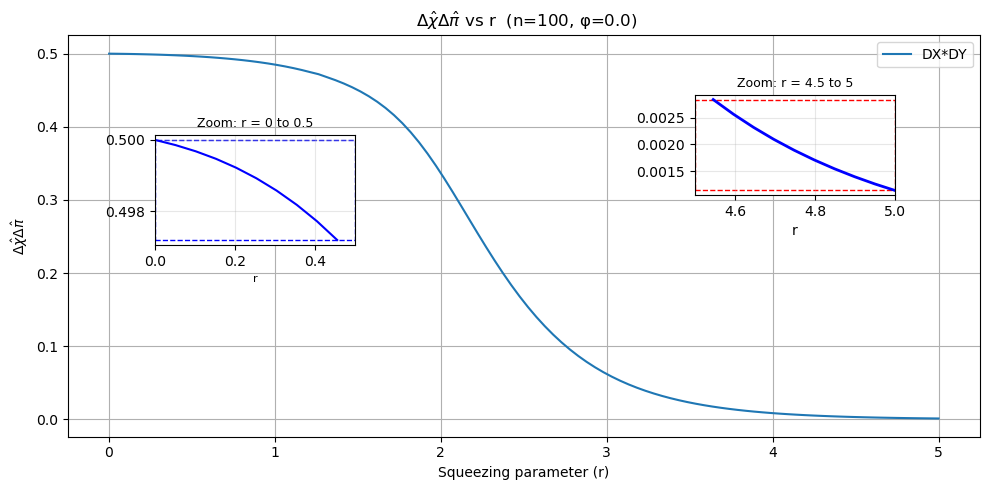}}
\caption{$\Delta\hat{\chi}\Delta\hat{\pi}$ calculated for the eigenstate of the squeeze operator indexed by $N_s=n$, the number of pairs, and different values of $r$. The insets show the behavior near the limits, $1/2$ and zero, that are dictated by the finite $N_s.$ }
\label{fig:framed_plot3}
\end{figure}

The product of the variances $\Delta\hat{\chi}\Delta\hat{\pi}$ is shown in Fig.~\ref{fig:framed_plot3}. As anticipated from the analytical discussion in Sect.~\ref{sec:properties}, the uncertainty product coincides with $(1-d)/2$, where $d=\langle D\rangle$ quantifies the deviation from the canonical commutation relations. Numerically, this result is obtained both from the direct evaluation of the variances and from the expectation value of the commutator $\langle [B,B^\dagger] \rangle = x^\dagger [B,B^\dagger] x$. The figure illustrates how the uncertainty product interpolates between the canonical bosonic limit and smaller values dictated by the finite number of fermion pairs. Importantly, this reduction does not signal any violation of uncertainty principles, but rather reflects the composite nature of the Frenkel-like cobosons and the associated state-dependent Heisenberg-Robertson bound.

\section{Concluding remarks} \label{sec:conclusion}

In this work, we have constructed and analyzed squeezed states for composite bosons made up of two spin-$1/2$ fermions, focusing on the case of Frenkel-like cobosons. Their internal fermionic structure leads to commutation relations that deviate from the cononical ones, which modify the usual properties of bosonic squeezing. By defining squeezed coboson states as eigenstates of a Bogoliubov-transformed coboson operator, we derived explicit expressions for the quadrature variances and showed how Pauli-blocking effects manifest themselves through a state-dependent Heisenberg–Robertson bound. In this context, the bound may fall below the canonical bosonic limit without implying any violation of the uncertainty principle. 

Our analytical results show that, while the formal structure of squeezing closely parallels that of elementary bosonic modes, the presence of a finite number of fermion pairs introduces quantitative deviations controlled by the expectation value of the operator $D.$ These deviations become particularly pronounced at large squeezing parameters and are consistently captured by our numerical analysis based on finite-dimensional matrix representations of the coboson quadratures.
The modified uncertainty relations derived here may have direct implications for observable quadrature fluctuations in systems of composite bosons. The reduction of the Heisenberg–Robertson bound originates from measurable deviations of the commutator expectation value 
$\langle [B,B^\dagger] \rangle$, which encode both Pauli blocking and finite pair occupancy. Consequently, squeezing in coboson systems manifests itself not only through the redistribution of fluctuations between conjugate quadratures, but also through an overall suppression of quantum noise that is intrinsically limited by compositeness. This behavior implies that experimentally accessible quantities, such as quadrature variances or noise spectra, carry information about the internal fermionic structure of the composite (quasi-)particles. In platforms such as exciton polaritons or other composite boson media, deviations from canonical bosonic squeezing therefore provide an operational signature of many-body correlations, rather than an anomaly of the uncertainty principle.

Although it goes beyond the scope of our paper to propose experimental implementations, we find that our study of squeezed Frenkel-like cobosons is more than a theoretical exercise. These objects and the approach we follow may provide a path to the quest of harnessing quantum noise reduction in condensed matter systems. The study of Frenkel excitons has the potential to provide schemes for quantum information processing, and a deeper understanding of quantum many-body physics, with potential applications ranging from fundamental biology to next-generation computing (even if not directly related some of these ideas can be found in \cite{chávezcarlos2024quantumsensingkerrparametric, plumhof}).

We are currently extending our analysis to Wannier-like cobosons \cite{Combescot_coherent}, where the nonuniform Schmidt structure and spatial delocalization are expected to introduce additional qualitative features in the squeezing behavior.

\section*{Acknowledgments} \label{sec:acknowledgements}

The authors would like to thank the Brazilian agency CAPES, for financial support. IR also thanks CNPq for partial support through contract 311876/2021.


\clearpage
\bibliographystyle{apsrev4-1}
\bibliography{refs.bib}

\end{document}